\documentclass[preprint,12pt]{elsarticle}
\usepackage{graphicx} 
\usepackage{amssymb}
\usepackage{lineno}

\begin{document}

\begin{frontmatter}

\title{Improved time-interval analysis}

\author[ifa]{H.O.U. Fynbo}
\author[ifa]{K. Riisager\corref{cor1} }
\address[ifa]{Department of Physics and Astronomy, Aarhus University,
 DK--8000, Aarhus C, Denmark}
\cortext[cor1]{Corresponding author: kvr@phys.au.dk, telephone +45 87155624}

\begin{abstract}
  Several extensions of the halflife analysis method recently
  suggested by Horvat and Hardy are put forward. Goodness-of-fit
  testing is included, and the method is extended to cases where more
  information is available for each decay event which allows
  applications also for e.g.\ $\gamma$ decay data. The results are tested
  with Monte Carlo simulations and are applied to the decays of
  $^{64}$Cu and $^{56}$Mn.
\end{abstract}

\begin{keyword}
Time interval  \sep  Dead time \sep Statistical analysis \sep Maxmimum likelihood
\end{keyword}

\end{frontmatter}

\section{Introduction}
The inevitable occurence of dead time when counting events from a
random process needs to be taken into account during analysis. The two
traditional limits, nonextending and extending deadtime, are
mathematically convenient but do not seem to describe actual data, see
\cite{Mul73,Mul91} and references therein for details. Recently an analysis
method for extracting halflives from decay data was suggested by
Horvat and Hardy \cite{Hor13} in which this problem is circumvented by
imposing a software extendable dead time. The method was shown to give
a significant reduction of the systematic error on the extracted
halflife.

The Horvat and Hardy paper focussed on the case where beta particles
were measured so that all events enter in the analysis. We explore here 
extensions to other cases where more information is available
for each event, the example we shall use is detection of $\gamma$-rays
where the energy is also recorded. In this case one can often select a
subset of events that have a more favourable signal to noise ratio,
e.g.\ by gating on a prominent $\gamma$ ray in the decay. This extension
is needed if one wishes to use the method also for electron capture
decays, one example being the decay of $^7$Be \cite{Maz12}.
The extension will also be useful in other cases where the total count
rate is sufficient to make dead time a potential problem, but the
fraction of interesting events is a smaller part of the total.
We shall furthermore show how the method of Horvat and Hardy
can be easily modified to provide a goodness-of-fit test as well and
compare to a histogram analysis procedure.

\section{Statistical formalism}
We first specify the notation. Decays follow an exponential
distribution, $\lambda \exp(-\lambda t)$ being the normalized
function.  The mean lifetime $\tau$ and the halflife $t_{1/2}$ are
related to the decay constant $\lambda$ by $t_{1/2} = \tau \ln(2) =
\ln(2)/\lambda$.  The total rate of events, $\rho$, when a background
is present is
\begin{equation}  \label{eq:rho}
   \rho(t) = A \lambda \exp(-\lambda t) + B \,,
\end{equation}
where $B$ is the rate of background events and $A$ can be interpreted
as the total number of decay events that could be recorded in an ideal
experiment (i.e.\ without dead time and with the measurement
continuing until all decays have happened). Our notation differs from
the one of Horvat and Hardy that include $\lambda$ in the front
factor; our choice leads to fit parameters with smaller correlation
which gives smaller final relative uncertainties on $A$. The
expression for $\rho$ can easily be generalized to the case where
several activities with each their halflife is present in the sample.

When a subset of interesting events can be selected, one divides all
events into two types $j=1,2$ (this can again be generalized).
Both groups of events may have both ``true decays'' and background
events and will have different decay rates
\begin{equation}  \label{eq:rhoj}
   \rho_j(t) = A_j \lambda \exp(-\lambda t) + B_j \,,
\end{equation}
where $A=A_1+A_2$ and $B=B_1+B_2$. The decay constant $\lambda$ is of
course the same for the two types. We let type 1 denote the events
that have the best signal to background ratio, i.e.\ $A_1/B_1 > A/B$.

The analysis takes place on $N$ events 
measured from the same sample with decay times $t_i$,
$i=1,2,\ldots,N$. The limitation that the data must result from one
time series can be lifted by analyzing simultaneously several sets of
data with each their parameters $A^k$, $B^k$ (the decay constant
$\lambda$ still being the same and the index $k$ denoting the
different sets). The method becomes impractical if the number of
produced samples is too large.

\subsection{The Horvat and Hardy procedure}
We refer to \cite{Hor13} for a detailed description of the Horvat and
Hardy procedure and will here give a brief alternative derivation of
it. There are two steps in the procedure. In the first step the time series
is pruned by imposing a fixed extendable dead time $t_e$ after each
recorded event. All events falling within a dead time window are
removed and the ``live time'' $t_l(i)$ preceeding each surviving event
(the time that has passed since the end of the last preceeding time
window until event number $i$) is calculated. In the second step, the
probability of each surviving event is calculated and combined into a
likelihood function. Since the rate $\rho$ depends (slowly) on time the
appropriate probability density is that of an nonhomogenous Poission
process (see e.g.\ equation (2.2.23) in \cite{Cox66})
\begin{equation}  \label{eq:nonhP}
  f_i = \rho(t_i) \exp \left( - \int_{t_i-t_l(i)}^{t_i} \rho(t') \mathrm{d}t' \right) \,.
\end{equation}
The final likelihood is then the product
\begin{equation}
  L = \prod_{i=1}^N  f_i \,.
\end{equation}
(Horvat and Hardy derive the expression for $f_i$ and include a factor
in $L$ corresponding to the time between the last event and the end of
the measurement. This last factor must be included to avoid a bias in
the method, but the bias is of order $N^{-1}$ which in many situations
is negligible.) The log-likelihood is finally given by $-2 \ln L$.
Note that the time series from a given sample need not be
uninterrupted. The
data taking may therefore include shorter or longer breaks, e.g.\ for
file change or when different samples are measured with the same
set-up.

It is straight-forward to extend this result by forming a likelihood
ratio $L/L_0$. This gives the possibility of performing also
goodness-of-fit tests. To form $L_0$ we optimize the ``rate
parameter'' independently for each event. In the approximation where
the rate is taken as constant over the time interval $t_l(i)$ this is
easily shown to give $\rho_i = 1/t_l(i)$. The final log-likelihood is
therefore
\begin{equation}   \label{eq:DP}
  D_P = -2\ln(L/L_0) = -2 \sum_{i=1}^N \left(  \ln(t_l(i)\rho(t_i)) -
      \int_{t_i-t_l(i)}^{t_i} \rho(t') \mathrm{d}t'  \right)   - 2N  \,.
\end{equation}
If $\lambda t_l(i) \ll 1$ one may increase the numerical precision by
Taylor expanding in the integral
\begin{equation}  \label{eq:Taylor}
  \int_{t_i-t_l(i)}^{t_i} \rho(t') \mathrm{d}t'  \approx  \rho(t_i)t_l(i)
  + A\lambda \exp(-\lambda t_i) \lambda t_l(i)^2 /2 + \ldots
\end{equation}

One reason for inserting the factor $-2$ into $D_P$ is that the
resulting quantity in many cases asymptotically will be $\chi^2$
distributed \cite{Jam06}. However, in our case the result is different
as shown in \ref{sec:etap}: the value per degree of freedom
$\nu = N - N_f$ ($N_f$ being the number of fit parameters) is $2\gamma
\approx 1.15442132$ with a standard deviation of $\approx
1.606/\sqrt{N}$. (A $\chi^2$ distribution would give $1$ and
$\sqrt{2/N}$, respectively.)

\subsection{Including information on event type}  \label{sec:seq}
If dead time was absent and one could select a subset of events with
better signal to background ratio, one would simply restrict the
halflife analysis to this subset. However, taking the dead time into
account implies one must work with the total recorded data set. Rather
than attempting to use the time intervals between type 1 events
directly, a procedure that would be strongly entangled with the
original analysis, we focus on including the independent information
on the type of event.

By far the simplest way to do this is to return to the expression for
a Poisson process and use the appropriate rate, $\rho_1$ or $\rho_2$
when the event is of type 1 or 2, explicitly instead of $\rho$ as the
first factor in equation (\ref{eq:nonhP}) (the integral in the
exponential of course still contains the total rate $\rho$). This
corresponds to introducing a factor $\rho_1/\rho$ for each event of
type 1 and a factor $\rho_2/\rho$ for each event of type 2. The value
obtained for $D_P$ will therefore increase by an amount that depends
on the size of $\rho_1/\rho$ for each event. The details are given in
\ref{sec:etaseq} that also discusses a more involved procedure for
analyzing the information of the type of event.

If more than one activity is present in the sample, due to
contaminants or daughter decays, this needs to be included in the rate
$\rho$. One may extend the above analysis to more than two types
simply by employing the appropriate decay rates for each type.
However, it is clearly possible even in the general case to work with
just two types, the interesting ones and ``the rest'', as long as
$\rho$ gives a satisfactory account of the event rate.

\subsection{Comparison to histogram analyses}
The need to include in the analysis all activities present in the
sample may at first glance seem to be a drawback of the method by
comparison to a histogram analysis with a selective gate.  The latter
will in favourable cases include only one activity and a background
term. However, if dead time is of concern one must also worry about
possible pile-up that could distort the histogram analysis since not
all pile-up events can be identified and rejected. Pile-up will remove
events from the selected gate and lead to a distortion of the decay
curve, a distortion that is much less pronounced if all events are
included, see \cite{Pom99,Pom99a} and references therein for a
detailed treatment of pile-up at high countrates.  Furthermore, many
$\gamma$ ray detectors will have a peak-to-total ratio much less than
one so that the counting statistics, and therefore the final
precision, is increased significantly if events outside the peak
region are included as well. Note that pile-up also will affect our
method in the previous subsection. In such cases one can use $\rho =
\rho_1 + \rho_2$ for all events in the start of the data set and
switch to using $\rho_1$ and $\rho_2$ when total count rates have
decreased to make pile-up negligible. The optimum analysis method will
depend on the exact experimental circumstances and it may pay off to
perform several analyses to check for systematic effects.

The time interval analysis is much slower than a conventional
histogram analysis and may become impractical for very large data
samples. An alternative procedure is then to remove events from the
recorded sequence that falls within the extendable dead time $t_e$, as
done above, but then project the resulting time sequence into a
histogram with appropriate time bin width. The histogram coming out of
this \emph{hybrid method} can then be analyzed with standard methods,
as given e.g.\ in equation (5) in \cite{Mul91}, since the dead time
now is known exactly. This procedure is significantly faster and
examples will be shown below.

Whatever analysis procedure is used one will in practice need to check
for contaminants in the data. One way is to do a visual inspection of
the (scaled) residuals between data and fit, as done in
\cite{Hor13}. Our results allow also to use the goodness-of-fit value
for a quantitative test. If need be, the goodness-of-fit value per
data point can also be used in a differential manner as check of
systematic deviations (giving a worse fit) for smaller sections of the
total data set. If the goodness-of-fit value had been
$\chi^2$-distributed one could also have made a quantitative analysis
based on the residuals, since the sum of the square of the scaled
residuals gives the $\chi^2$.

\section{Illustration of the method}
\begin{figure}
\centering
  \includegraphics[width=.99\textwidth]{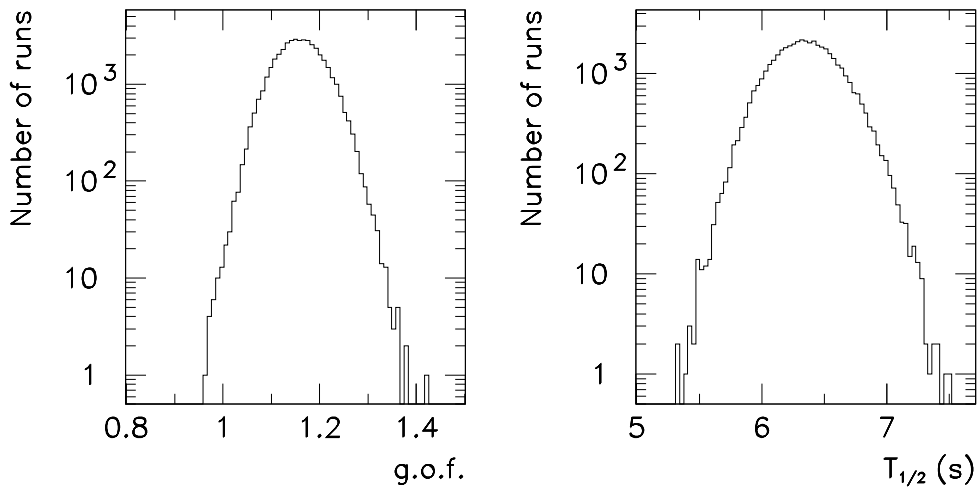} 
  \caption{The result of analysis of 43695 Monte Carlo simulated
    samples with halflife 6.3452 s and an initial event rate of
    100/s. The left panel shows the obtained values of $D_P$ and the
    right panel the extracted halflife values. Both have their expected
  mean value and a spread consitent with the statistics.}
\label{fig:MC100}   
\end{figure}

The methods developed above will first be tested on Monte Carlo
data. For the original $D_P$ (without goodness-of-fit) much of this
was done in \cite{Hor13} and we will only add a few results to what
was obtained there. Figure \ref{fig:MC100} gives results of
simulations performed with similar parameters as in \cite{Hor13},
namely a halflife of 6.3452 s, a background rate of 1/s and an initial
activity of 100/s. A software extendable dead time of 2 $\mu$s is
used and about 1000 decays are recorded per run. The analysis is seen
to reproduce well the input parameters (up to terms of order $N^{-1}$)
and the goodness-of-fit value is seen to be centered around $2\gamma$,
as expected.  Simulations have also been carried out for an initial
activity of 1000/s and give similar good agreement. See
\ref{sec:halflife} for general comments on the uncertainty on the
halflife.

The hybrid method with projection into histograms with bin width 1s
gave in this case the same results as the $D_P$ analysis. (Here the
Poisson likelihood $\chi^2_{\lambda}$ was used that, similar to $D_P$,
does not give a goodness-of-fit per channel of one for low count
numbers, see \cite{Ber02} for details.) For high initial counting
rates a difference between a time interval analysis and a histogram
analysis was found in \cite{Hor13}. The histograms were there
projected out using a non-extendable dead time, whereas we use an
extendable dead time. In our simulations we do not find as strong
differences between the two methods as reported in \cite{Hor13}, but
great care must be taken when applying the corrections for an
extendable dead time when the corrections become sizeable (for values
of the product of rate and dead time above 0.02 \cite{Hor13}), so we
also recommend that the $D_P$ analysis is used whenever the initial
counting rates are high.

If the uncertainty on the rate parameter is of interest (e.g.\ in
cross-section determinations in activation experiments), it is
important to use the parametrisation in equation (\ref{eq:rho}) rather
than including $\lambda$ in the factor in front of the exponential. In
the latter case the correlation coefficient with the halflife will
always be large: the final derived uncertainty on the parameter $A$ is
in the simulations found to be more than 40\% larger.

\begin{figure}[tbh]
\centering
  \includegraphics[width=.8\textwidth]{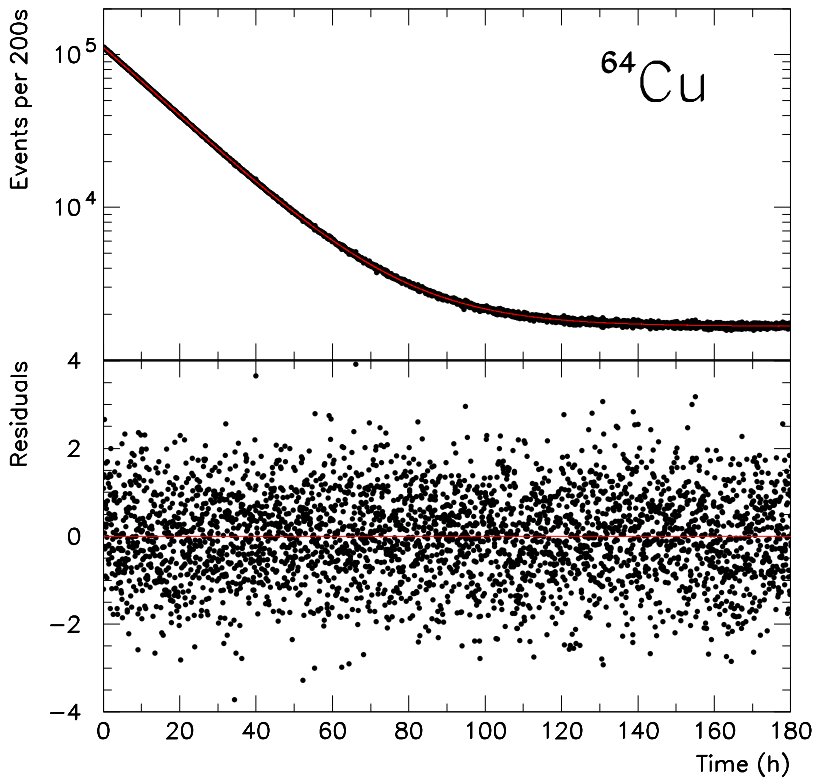} 
  \caption{The top panel shows the recorded number of decay events for
    $^{64}$Cu per 200 seconds versus time. The red line through the
    points is the fit to the data. The bottom panel shows the
    corresponding residuals, (data-fit)/uncertainty.}
\label{fig:Cu64}   
\end{figure}

The method was employed on the nuclei $^{64}$Cu and $^{56}$Mn
produced via neutron activitation of Cu and Mn foils. The literature
values for the halflife of the two nuclei are 12.7004(20) h
\cite{Be12} and 2.5789(1) h \cite{Jun11}, respectively. The Cu sample
will contain the $^{66}$Cu 5.1 min activity, and the data from the
first hour is therefore excluded from the spectrum shown in figure
\ref{fig:Cu64}. The activity coming from the decays, the 511 keV
annihilation $\gamma$ ray for $^{64}$Cu and several $\gamma$ rays for
$^{56}$Mn, were detected with a standard Ge detector.  
In the analysis a software extendable time interval of 100 $\mu$s was
used.

For both nuclei the final error on the halflife decreases by about a
factor two when all data are used rather than the restricted
gate. Furthermore, analysis of histograms for gated events gives a too
high halflife in both cases, most probably due to effects of
pile-up. We therefore quote only results corresponding to analysis of
all $\gamma$-ray events and check the effect of delaying the starting
point of the analysis to make sure that short-lived contaminants are
not included. The final extracted halflives are 12.710(3) h for
$^{64}$Cu and 2.5800(7) h for $^{56}$Mn, where the initial count rates
were 585/s and 1310/s. The results of the time
interval analysis and the hybrid method agree, with a tendency for the
former to be more robust (the histogram results for $^{56}$Mn vary
more).  In both cases our values are higher than the current
literature values, although only a few (of our) standard
deviations. It may be that some of the previous high-precision
experiments with initial high counting rates were prone to the
systematic effect pointed out by Horvat and Hardy \cite{Hor13}. The
residuals for both Mn and Cu (bottom panel in figure \ref{fig:Cu64})
give no indications for remaining systematic effects. The
goodness-of-fit value per degree of freedom is in both cases slightly
above the theoretical one, $54604953/47194135 = 1.1570$ for $^{64}$Cu
and $23787149/20469106 = 1.1621$ for $^{56}$Mn (expected deviations
from 1.1544 are 2 and 4 on the last digit, respectively). This was
traced to a periodic structure (period around 400 $\mu$s) in the time
interval spectra that must be due to the data acquisition system, but
is not expected to influence the extracted halflife values. The
$\chi^2$ values from the corresponding histogram analyses are
acceptable (3354/3228 and 2543/2418).

Including information on the event type will always lead to a
reduction of the uncertainty on the extracted halflife, but the
simulations show that this reduction is only significant when the
initial source and background activities are comparable. As an
example, with the same parameters as above of halflife 6.3452 s and
initial activity 100/s but background rate increased to 100/s one gets
an uncertainty of 0.76 s for the extracted halflife; if a type 1 event
in this case has initial activity 25/s and background 1/s the
uncertainty decreases to 0.53 s (with an increase of gof/dof of
0.22). Even though only a quarter of the decay events are of type 1
(therefore giving twice the uncertainty if no background had been
present) the signal to background ratio is sufficiently more
favourable to give a significant reduction in final uncertainty.  As a
realistic example, we have repeated the analysis of the $^{64}$Cu
spectrum including only data after 72 h where Cu and background
activities in the full spectrum are about equal. The time interval
analysis of all data gives here a halflife of 12.674(49) h with a
gof/dof of $4981984/4309914 = 1.1559$ (expected spread 8 on the last
digit) whereas analysis with a gate on the 511 keV line gives a
halflife of 12.702(42) h and a gof value of 7031447 (the increase per
dof of 0.488 is consistent with the results given in
\ref{sec:etaseq}). The increase in precision is small, but
significant.  If the time interval method is employed for experiments
with charged particle detection rather than $\gamma$ ray detection,
the ``good events'' can be much more concentrated in energy and it
will always be advantageous to use the information on event type.

\section{Conclusions}
The basis for all the analysis performed here is the assumption that
effects of dead time will have a maximum extent $t_e$ in time ($t_e$
can be varied to achieve this): for shorter time intervals $\Delta t$
between events there may be corrections to the usual exponential
waiting time distribution, but for larger intervals we can safely
assume the exponential distribution. Only events with $\Delta t > t_e$
are selected. In the time interval analysis $t_l = \Delta t - t_e$ is
used for further analysis via equation (\ref{eq:DP}), whereas the data
set are projected into a time histogram in the hybrid method. If more
information is available for the event, e.g.\ whether a $\gamma$-ray
energy falls within a ``good window'' or not, this may be used to
refine the analysis. In both cases one needs to evaluate the effects
of dead time from the full data set, so that e.g.\ a correction found
from the histogram based on all events can be used for the histogram
with the gated spectrum.

The results of the simulations and analysis of real data presented in
this paper can be summarized in the following recommendations.
If there are only moderate dead time effects one may expect pile-up to be
negligible. In this case it should be safe to do analysis dividing the
data into several types (a gate in energy etc), but if doable one
should always check results with the full data set. If the count rates
are high analysis should only be done on the full data set.
Concerning the analysis method $D_P$ should be used unless
the amount of data is very large and the product of the initial
counting rate and the imposed dead time $t_e$ is less than 0.01.

Our normalization of $D_P$ allows to use it for goodness-of-fit
tests. It can be a very sensitive indicator for the presence of
contaminants in the data. The small spread of the expected value of
$D_P$, caused by the large value of $N$, is not kept when more
information is included, but figure \ref{fig:etaseq} still allows to
test goodness-of-fit on about the same level as a $\chi^2$ test in the
histogram analysis.

The methods were employed to analyse decay data from two
neutron-activated samples and gave halflives of 12.710(3) h for
$^{64}$Cu and 2.5800(7) h for $^{56}$Mn. These values are free from
pile-up and dead time effects and it is noteworthy that the precision
obtained here from a single run is approaching the one reached in many
of the previous halflife determinations for these two isotopes.

The methods put forward here allow the time interval analysis to be
applied to a larger range of problems than beta counting including
halflife determinations in more exotic nuclear decays \cite{Pfu12}.

\section*{Acknowldegements}
We would like to thank Jens Ledet Jensen for valuable advice on
statistical issues.

\appendix
\section{Statistical details}

There is an extensive statistical litterature on Possion processes,
see e.g.\ \cite{Cox66} that gives a detailed overview of the many
tests that have been employed. Due to the dead time we do not record
all decay events, the changes this gives for a standard histogram
analysis are described e.g.\ in \cite{Mul73}.  We here follow
\cite{Hor13} and analyse instead the distribution of times between
individual recorded events that furthermore is ``truncated from the
left'' by the imposed software dead time $t_e$. Since there is no
memory in a decaying system, we still have an exponential distribution
for each time interval, leading in the general case to the expression
in equation (\ref{eq:nonhP}).

\subsection{Uncertainty on halflife}  \label{sec:halflife}
If no background is present it is easy to show from
maximum likelihood (see e.g.\ \cite{Bar89}) that the mean of the
recorded decay times $\bar{t}$ is the best fit value for $\tau$ and
that the relative uncertainties become $\sigma(t_{1/2})/t_{1/2} =
\sigma(\tau)/\tau = 1/\sqrt{N}$. The case where only events up to a
time $T$ are recorded can also be treated analytically and yields a
relative uncertainty of
\begin{equation}
   \frac{\sigma(t_{1/2})}{t_{1/2}} = \frac{\sigma(\tau)}{\tau} = 
    \frac{1}{\sqrt{N}} \left[ 1 - \left( \frac{T}{\tau} \right)^2
  \frac{ \exp(T/\tau)}{(\exp(T/\tau)-1)^2} \right]^{-1/2}  \;. 
\end{equation}
This shows explicitly the well-known fact that decays should be
followed for many halflives in order to extract a precise value. The
presence of a background limits how far in time it is meaningful to
follow a decay, but having the possibility to gate on a subset of
events may extend this limit.

It may be illustrative to trace how the final uncertainty on the
halflife emerges in our case. This can be done by using
\cite{Jam06,Bar89} that the variance on the parameter $\lambda$ can be
estimated as the inverse of the expectation value of:
\begin{equation}
   - \frac{\partial^2 \ln L}{\partial \lambda^2} =
     \sum_{i=1}^N \frac{1+t_i(2-\lambda t_i)B/(Ae^{-\lambda t_i})}
      {(\lambda + B/(Ae^{-\lambda t_i}))^2}   +  A \sum_{i=1}^N 
      \left( t_i^2 e^{-\lambda t_i} - (t_i-t_l(i))^2 e^{-\lambda (t_i-t_l(i))} \right) \;,
\end{equation}
where the simplest case of equation (\ref{eq:rho}) was used. Here the
second term quantifies the effect of the dead time: if no dead time is
present one has $t_i-t_l(i) = t_{i-1}$ and the whole term gives
zero. If no background is present, the first term gives $N/\lambda^2$
and therefore the standard uncertainty of $\lambda/\sqrt{N}$. With a
background term the contributions to the sum decrease when $B$ becomes
larger than $A\lambda e^{-\lambda t_i}$ which will increase the final
error on $\lambda$. The effective cut off time for how long it makes
sense to continue a measurement is therefore of order the time where
the source activity and background activity are equal. Continuing data
taking will reduce the uncertainty on the background rate.

\subsection{Goodness of fit for $D_P$} \label{sec:etap} 
The quantity $-2$ times the log-likelihood ratio is sometimes called
deviance in the statistical literature \cite{Nel72} and we therefore use the name
$D_P$ for our case where it is applied to a Poisson process. We
generalize the case from the main text slightly and assume that the
decay rate can be written as a sum of terms where the amplitudes are
independent (there may be more variables $b$, typically decay rates,
in each term and they need not be independent)
\begin{equation}
   \rho(t) = \sum_{l=1}^M a_l \rho_l(b,t)  \,,
\end{equation}
so that $\partial \rho/\partial a_l = \rho_l(b,t)$. As an example, in the case
considered in the main text at equation (\ref{eq:rhoj}) one has
$a_1 = A_1$, $a_2 = B_1$, $a_3 = A_2$ and $a_4 = B_2$.
Then at the minimum of $D_P$ we have
\begin{equation}   \label{eq:deriv_al}
   0 = \frac{\partial D_P}{\partial a_l} =
   -2 \sum_{i=1}^N \frac{\rho_l(b,t_i)}{\rho(t_i)} +
   2 \sum_{i=1}^N \int_{t_i-t_l(i)}^{t_i} \rho_l(b,t') \mathrm{d}t'  \,.
\end{equation}
Inserting these relations in $D_P$ gives
\begin{eqnarray}
   D_{P,min} & = & -2\sum_{i=1}^N \ln(\hat\rho t_l(i)) + 2\sum_{i=1}^N
   \int_{t_i-t_l(i)}^{t_i} \sum_{l=1}^M a_l \hat\rho_l(b,t') \mathrm{d}t' - 2N \\
   & = & -2\sum_{i=1}^N \ln(\hat\rho t_l(i)) + 2 \sum_{l=1}^M a_l \sum_{i=1}^N
   \frac{\hat\rho_l(b,t_i)}{\hat\rho(t_i)}- 2N =  -2\sum_{i=1}^N \ln(\hat\rho t_l(i)) \nonumber 
\end{eqnarray}
since the middel term can be rewritten to $\sum_{i=1}^N 1 = N$. The
result actually holds \cite{Nel72} also for the more general case of a
gamma distribution (the exponential distribution being a special
case). The hat added to $\rho$ is to remind that it is the decay rate
with fit parameters inserted, the restrictions coming from
minimization with respect to the parameters $b$ (in our case the
halflife) also have to be added. For the corresponding case of a
$\chi^2$-distribution the main effect of these restrictions is to
reduce the number of degrees of freedom of the distribution (by up to
the number of parameters). We shall assume a similar behaviour here;
in general there may be correction terms of order $1/\sqrt{N}$ (or
higher orders), see \cite{McC89} for general asymptotic results on the
deviance for gamma distributions.

If the decay rate describes the data we have an exponential
distribution $\exp(-x)$ for all parameters $x_i = \rho(t_i)
t_l(i)$. In the limit where the decay rate can be taken as constant
over $t_l(i)$ each term in the sum in equation (\ref{eq:DP}) reduces
to $g(x) = 2(x-\ln x -1)$. The expectation value of this is $\langle g
\rangle = 2\gamma$, $\gamma$ being Euler's constant, and from $\langle
g^2 \rangle$ one obtains the variance of $4\pi^2/6-4$.  The way our
likelihood ratio is constructed we have (for $L_0$) as many parameters
as data points, so we are not in the asymptotic limit where
log-likelihood ratios become distributed as $\chi^2$
\cite{Jam06}. However, the central limit theorem can be used and gives
that the expected value of $D_P$ per degree of freedom becomes
$2\gamma$ with a standard deviation of $2\sqrt{\pi^2/6-1}/\sqrt{N}$.

\subsection{The event type analysis}   \label{sec:etaseq}
\begin{figure}[tbh]
\centering
  \includegraphics[width=.8\textwidth]{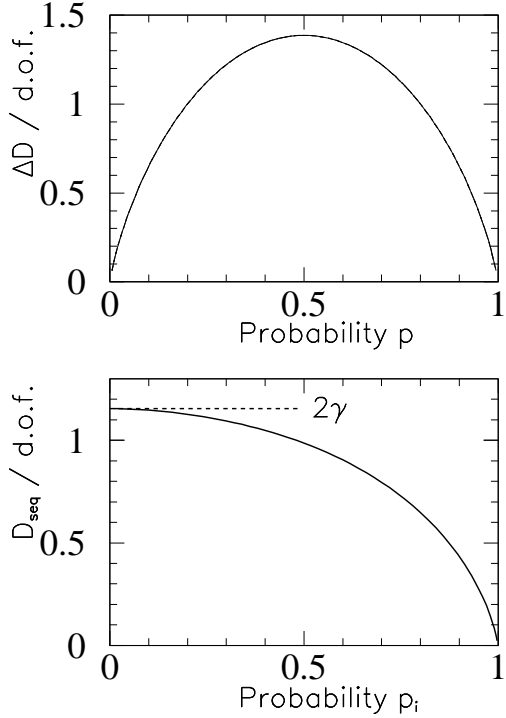} 
 \caption{Top panel: the extra contribution to $D_P$ per degree of
    freedom from the event type analysis is shown as a function of $p
    =\rho_1/\rho$.  Bottom panel: The contribution to $D_{seq}$ per
    degree of freedom, given in equation (\protect\ref{eq:etaseq}), is
    shown as a function of the probability $p_i$.}
\label{fig:etaseq} 
\end{figure}

In the procedure in section \ref{sec:seq} the likelihood $L$ was
modified to
\begin{equation}  
  L' = \prod_{i=1}^N \rho_{I_i}(t_i) \exp \left( - \int_{t_i-t_l(i)}^{t_i} \rho(t') \mathrm{d}t' \right) \,.
\end{equation}
where $I_i = 1$ or 2 according to the event type, whereas $L_0$ was
left unchanged. In the limit where all decay rates are slowly varying
one can interpret $p = \rho_1/\rho$ as the (local) probability of
obtaining an event of type 1, $1-p$ then being the probability of
obtaining type 2. Based on this one would expect
$-2\ln(\rho_{I_i}/\rho)$, the extra contribution to $D_P$ per degree
of freedom, to give $-2(p\ln p + (1-p) \ln (1-p))$, where the average
has to be taken if $p$ varies substantially during the data set. The
extra contribution is plotted in figure \ref{fig:etaseq} and goes, as
expected, to zero in the two limits of $p\rightarrow 0$ and
$p\rightarrow 1$ where one type of events dominate.

The same modification of $L$ can be derived in the following manner:
Let there be a total of $N_1$ events of type 1 and let $n_i$ be the
number of events of type 2 between the (i-1)'th and i'th event of type
1. As an example, for the sequence
\[  \ldots 121222211221 \ldots \]
the values of $n_i$ are 1, 4, 0 and 2.
We keep the original analysis of $L$ and include the information on
the type of event with a separate likelihood function for the recorded
sequence of types which becomes
\begin{equation}
  L_{seq} = \prod_{i=1}^{N_1} P_i\,, \; P_i = p_i(1-p_i)^{n_i} \,.
\end{equation}
(Note that this is a geometric distribution \cite{Jam06} in $n_i$ and
not a binomial distribution since the type 1 event by definition is at
the beginning of the i'th subsequence.)  In doing this we disregard
the possible events of type 2 that follow the last event of type 1.

It is easy to see that $LL_{seq}$ is the same as the modified $L'$
above, but the alternative derivation allows to form a likelihood ratio
also for $L_{seq}$ by optimizing each $p_i$ individually which
results in $p_i = 1/(n_i+1)$ and gives
\begin{equation}
   P_i^{ref} = \frac{n_i^{n_i}}{(n_i+1)^{n_i+1}}\,
\end{equation}
where $0^0=1$ must be used. The contribution $D_{seq} = -2 \ln
(L_{seq} / L_{seq}^{ref})$ to the log-likelihood from this
sequence-analysis therefore becomes
\begin{equation}
  D_{seq} =   -2 \sum_{i=1}^{N_1}
      [\ln p_i +n_i\ln(1-p_i) -n_i\ln n_i + (n_i+1)\ln(n_i+1))] \;.
\end{equation}
The corresponding number of degrees of freedom is $N_1-2$ since two
extra parameters ($A_1$, $B_1$) were introduced in the analysis. 

We should consider whether the terms in $D_{seq}$ changes the
analysis in \ref{sec:etap} and restrict, to simplify
notation, ourselves to the case in equation (\ref{eq:rhoj}).
In this approximation the extra terms that will be
added to equation (\ref{eq:deriv_al}) can be written as
\begin{equation}
  -\sum_{i=1}^{N_1} \left( 
    \frac{1}{\rho_1} \frac{\partial \rho_1}{\partial a_l}  -
    \frac{1}{\rho} \frac{\partial \rho}{\partial a_l}   \right)
  -\sum_{i=1}^{N_1}  n_i \left( 
    \frac{1}{\rho_2} \frac{\partial \rho_2}{\partial a_l}  -
    \frac{1}{\rho} \frac{\partial \rho}{\partial a_l}   \right) \;.
\end{equation}
Explicit evaluation of the four different terms gives a combined
contribution to $D_{P,min}$ of
\begin{equation}
  \sum_{i=1}^{N_1} \left[ \rho_1 (\frac{1}{\rho_1}-\frac{n_i+1}{\rho})
     + \rho_2 (\frac{n_i}{\rho_1}-\frac{n_i+1}{\rho})  \right] = 0 \;,
\end{equation}
so that the addition of $D_{seq}$ will not change the expected
value for the $D_P$ part (assuming that the total decay rate
$\hat\rho$ is unchanged).

Turning then to the contribution from $D_{seq,min}$ we first
calculate the expectation value of the i'th term $\ln(P_i/P_i^{ref})$
that is
\begin{equation}
  \ln p_i + \ln (1-p_i) \sum_{n=o}^{\infty} n p_i (1-p_i)^n - 
     \sum_{n=0}^{\infty} p_i(1-p_i)^n \ln\frac{n^n}{(n+1)^{n+1}} \;,
\end{equation}
which can be rewritten as
\begin{equation}  \label{eq:etaseq}
  \ln p_i + \frac{1-p_i}{p_i}\ln(1-p_i) +
    \frac{p_i^2}{1-p_i}  \sum_{n=2}^{\infty} (1-p_i)^n n \ln n
\end{equation}
that has to be evaluated numerically. The result is displayed in
figure \ref{fig:etaseq}. As can be seen from the expression the
contribution goes to zero as $p_i$ approaches one; this is the limit
where all events are of type 1 and the extra analysis therefore does
not add any information. In the opposite limit where $p_i$ goes to
zero the contribution seems numerically to approach $2\gamma$.
Note that since $N_1 = p N$ the two first terms in equation
(\ref{eq:etaseq}) gives the same overall contribution as the extra
contribution to $L'$.

\end{document}